\documentclass[aps,pra,showpacs,twocolumn]{revtex4-1}
\usepackage{amssymb}
\usepackage{amsmath}
\usepackage{graphicx}
\usepackage{epsfig}

\begin{document}

\title{Quantum mold casting for topological insulating and edge states}
\author{X. M. Yang and Z. Song}
\email{songtc@nankai.edu.cn}
\affiliation{School of Physics, Nankai University, Tianjin 300071,
China}
\begin{abstract}
We study the possibility of transferring fermions from a trivial system as
particle source to an empty system but at topological phase as a mold for
casting a stable topological insulator dynamically. We show that this can be
realized by a non-Hermitian unidirectional hopping, which connects a central
system at topological phase and a trivial flat-band system with a periodic
driving chemical potential, which scans over the valence band of the central
system. The near exceptional-point dynamics allows a unidirectional
dynamical process: the time evolution from an initial state with full-filled
source system to a stable topological insulating state approximately. The
result is demonstrated numerically by a source-assistant QWZ model and SSH
chain in the presence of random perturbation. Our finding reveals a
classical analogy of quench dynamics in quantum matter and provides a way
for topological quantum state engineering.
\end{abstract}

\maketitle

\section{Introduction}

Preparing a topologically nontrivial state is of great interest for the task
of quantum information processing. In general, a natural way for preparing a
topological insulating state is cooling a system down to its ground state by
suppressing the thermal fluctuations.\ An other intuitive way for
preparation of topological insulating state is to follow a mechanical way by
filling the topological energy band with fermions dynamically. However, it
is tough to move fermions one by one from a particle source to an empty
system since the fermion obeys the Schrodinger equation rather than the
Newton's laws. Recently, a dynamical way of realizing the topological phases
by applying a time periodic global driving on a topologically trivial
initial state is proposed. It has been shown that these periodic
perturbations lead to the realization of new topological phases of matter
which have no equilibrium counterparts \cite{TOKA, TKIT, MSRU, LEFF, HDEH,
JHWI}, including topological insulators \cite{NHLI, JCAY} and edge states 
\cite{MTHA, AKUN}. It opens up a way to realize a topological phase
dynamically. In these studies above, both the effective Hamiltonian
dictating the non-equilibrium dynamics of the system and the initial static
Hamiltonian are Hermitian. Nevertheless, a non-Hermitian Hamiltonian is no
longer forbidden both in theory and experiment since the discovery that a
certain class of non-Hermitian Hamiltonians could exhibit entirely real
spectra \cite{Bender,Bender1}. The origin of the reality of the spectrum of
a non-Hermitian Hamiltonian is the pseudo-Hermiticity of the Hamiltonian
operator \cite{Ali1}. It motives a non-Hermitian extension of the dynamical
preparation of a topologically\ nontrivial state. In addition, the peculiar
features of a non-Hermitian system do not only manifest in statics but also
dynamics. Non-Hermitian systems exhibit many peculiar dynamic behaviors that
never occurred in Hermitian systems. One of the remarkable features is the
dynamics at exceptional point (EP) \cite{HeissEP,RotterEPJPA,HXu} or
spectral singularity (SS) \cite{AMprl,SLHprb,AAA,BFS,ZXZ2}, where the system
has a coalescence state. Exclusively, EP dynamics has recently emerged as
transformative tools for dynamically evolving quantum systems into a quantum
phase with desirable properties \cite{XZZH, XZZHarxiv, XMYA}. It is expected
that the introduction of non-Hermitian elements benefits to the scheme for
quantum state engineering.

In this work, we focus on the EP-related dynamic behavior for the many-body
system.\ From the perspective of non-Hermitian quantum mechanics, it is also
a challenge to deal with many-particle dynamics. As an application, we study
the possibility of transferring fermions from a trivial system as particle
source to an empty system but at topological phase as a mold for casting a
stable topological insulator dynamically. We show that this can be realized
by a non-Hermitian connection between a central system at topological phase
and a flat-band system with a periodic driving chemical potential. After
sufficient long time, the near exceptional-point dynamics allows the time
evolution from an initial state with full-filled source system to a stable
topological insulating state approximately. We demonstrate the scheme by
numerical simulations for a source-assistant QWZ model and SSH chain in the
presence of random perturbation. The result reveals a classical analogy of
quench dynamics in quantum matter and provides a way for topological quantum
state engineering. It also shows that a unidirectional tunneling supports an
exclusive feature never occurs in Hermitian system, which can be utilized to
quantum state engineering. Our findings pave the way for establishing
EP-dynamics based techniques as a powerful and versatile tool for
topological state engineering.

This paper is organized as follows. In Section \ref{Model and coalescing
state}, we describe the model Hamiltonian and give the condition for the
existence of coalescing state. In Section \ref{EP dynamics and periodic
driving}, based on the solutions, we present the characteristics of the EP
dynamics in a source-assistant QWZ model and the details of dynamical
formation of topological insulating state. In Section \ref{Edge states
engineering}, we propose a dynamical way to cast edge states in a
source-assistant RM model.\ Finally, we give a summary and discussion in
Section \ref{sec_summary}.

\begin{figure*}[tb]
\centering
\includegraphics[ bb=5 398 570 775, width=14 cm, clip]{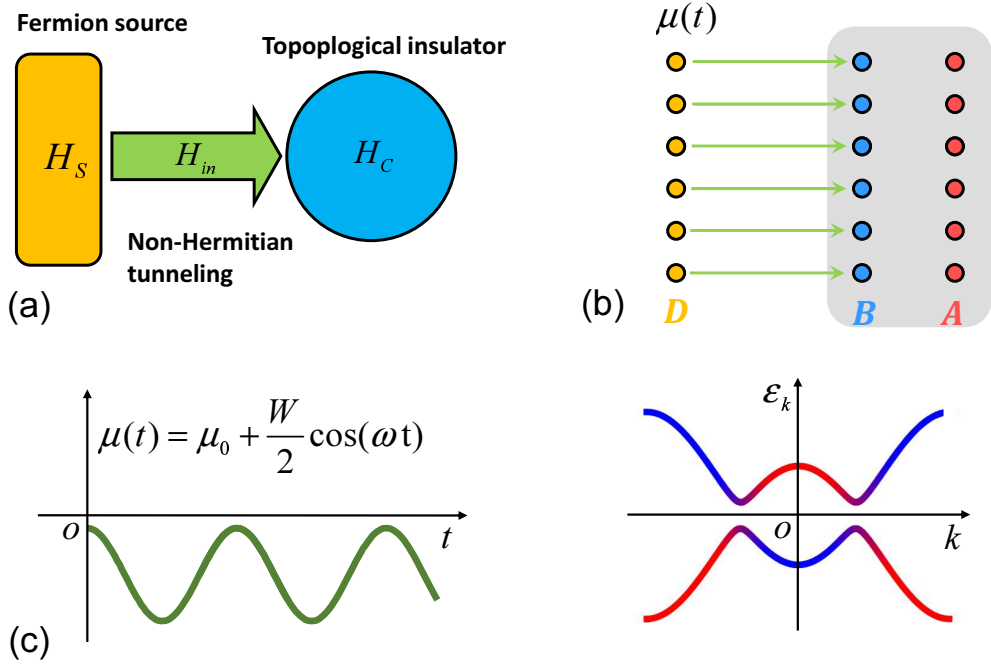}
\caption{ Schematics for the system and process of quantum mold casting. (a)
The system consists of two parts, central system $H_{\mathrm{c}}$\ and
source system $H_{\mathrm{s}}$. The target state is the ground state of $H_{%
\mathrm{c}}$, which can be topologically non-trivial or not. $H_{\mathrm{s}}$%
\ is a topologically trivial\ system, providing the supply of fermions.\
Both $H_{\mathrm{c}}$\ and $H_{\mathrm{s}}$ are Hermitian, while $H_{\mathrm{%
in}}$\ is non-Hermitian,\ representing the connection between\ $H_{\mathrm{c}%
}$\ and $H_{\mathrm{s}}$, and taking the role to transport fermions
unidirectionally from $H_{\mathrm{s}}$ to\ $H_{\mathrm{c}}$.\ (b) A
tight-binding model for the scheme, which contains three sets, A, B, and D.
Lattices A and B (red and blue filled circles) embedded in the shadow area
is topological insulator, while lattice D (yellow filled circle) is a
flat-band (hopping-free) system but with oscillating chemical potential.
Green arrows represent unidirectional hopping from\ D to B lattices. The aim
of this work is to realize the following process via time evolution.
Initially, D lattice is fully filled, while A and B are empty. The final
state is expected to a half-filled ground state of $H_{\mathrm{c}}$. (c) The
underlying mechanism of the dynamic process. At instant $t_{\mathbf{k}}$,
the chemical potential and energy levels of $H_{\mathrm{c}}$\ are resonant,
leading to exceptional points. The corresponding (EP) dynamics allows a
complete transfer of fermions between the degenerate energy levels.\ In the
long-time limit, such dynamics occurs at each $\mathbf{k}$\ sector again and
again. It is expected the lower band of $H_{\mathrm{c}}$\ can be completely
filled.}
\label{fig1}
\end{figure*}

\section{Model and coalescing state}

\label{Model and coalescing state}

We consider a non-Hermitian time-dependent Hamiltonian%
\begin{equation}
H=H_{\mathrm{c}}+H_{\mathrm{s}}+H_{\mathrm{in}},
\end{equation}%
with%
\begin{equation}
H_{\mathrm{c}}=\sum_{i,j=1}^{N}\left( T_{ij}a_{i}^{\dag
}b_{j}+A_{ij}a_{i}^{\dag }a_{j}+B_{ij}b_{i}^{\dag }b_{j}+\mathrm{H.c.}%
\right) ,
\end{equation}%
and%
\begin{equation}
H_{\mathrm{s}}=\mu (t)\sum_{j=1}^{N}d_{j}^{\dag }d_{j}\text{, }H_{\mathrm{in}%
}=\gamma \sum_{j=1}^{N}b_{j}^{\dag }d_{j},
\end{equation}%
where $a_{j}$, $b_{j}$,\ and $d_{j}$\ are fermion operators and $\mu (t)=\mu
_{0}+\frac{W}{2}\cos (\omega t)$\ is periodic driving chemical potential,
with parameters $\left\{ T_{ij},A_{ij},B_{ij}\right\} $\ depending on $H_{%
\mathrm{c}}$ ($\mu _{0}$\ is an average of the negative energy levels and $W$
is the bandwidth of $H_{\mathrm{c}}$). Here both $H_{\mathrm{c}}$\ and $H_{%
\mathrm{s}}$ are Hermitian, describing the central system and source system,
respectively. Notably, $H_{\mathrm{in}}$\ is a non-Hermitian term,\
representing the connection between two systems\ $H_{\mathrm{c}}$\ and $H_{%
\mathrm{s}}$.

For a central system with the periodic boundary condition, after performing
Fourier transformation, we have $H=\sum_{\mathbf{k}}H_{\mathbf{k}}$, where
the sub-Hamiltonian $H_{\mathbf{k}}$ in each invariant sub-space can be
expressed as%
\begin{equation}
H_{\mathbf{k}}=\left( 
\begin{array}{ccc}
a_{\mathbf{k}}^{\dag } & b_{\mathbf{k}}^{\dag } & d_{\mathbf{k}}^{\dag }%
\end{array}%
\right) h_{\mathbf{k}}\left( 
\begin{array}{c}
a_{\mathbf{k}} \\ 
b_{\mathbf{k}} \\ 
d_{\mathbf{k}}%
\end{array}%
\right) ,
\end{equation}%
where%
\begin{equation}
\left( 
\begin{array}{ccc}
a_{\mathbf{k}}^{\dag } & b_{\mathbf{k}}^{\dag } & d_{\mathbf{k}}^{\dag }%
\end{array}%
\right) =\sum_{\mathbf{r}}\frac{e^{i\mathbf{k\cdot r}}}{\sqrt{N}}\left( 
\begin{array}{ccc}
a_{\mathbf{r}}^{\dag } & b_{\mathbf{r}}^{\dag } & d_{\mathbf{r}}^{\dag }%
\end{array}%
\right) .
\end{equation}%
In general, the Bloch matrix $h_{\mathbf{k}}$\ has the form

\begin{equation}
h_{\mathbf{k}}=\left( 
\begin{array}{ccc}
B_{z}(\mathbf{k)} & B_{x}(\mathbf{k)}-iB_{y}(\mathbf{k)} & 0 \\ 
B_{x}(\mathbf{k)}+iB_{y}(\mathbf{k)} & -B_{z}(\mathbf{k)} & \gamma \\ 
0 & 0 & \mu (t)%
\end{array}%
\right) ,
\end{equation}%
where a term related to the identity matrix $I_{3}$\ is neglected. Here,
vector $\mathbf{B}(\mathbf{k)}$\ is obtained by the set of parameters $%
\left\{ T_{ij},A_{ij},B_{ij}\right\} $. We note that the time-dependent $\mu
(t)$\ does not break the translational symmetry (also other all symmetries),
allowing the exact diagonalization of $H$\ via that of each $3\times 3$\
matrix. Accordingly, the time evolution can also be computed via the
complete set of $3\times 3$\ matrices.

Here we focus on two points of particular interest: i) We note that in $k$
space, the central Hamiltonian can be expressed as 
\begin{equation}
H_{\mathrm{c}}=\sum_{\mathbf{k}}\left( 
\begin{array}{cc}
a_{\mathbf{k}}^{\dag } & b_{\mathbf{k}}^{\dag }%
\end{array}%
\right) \mathbf{B}_{\mathbf{k}}\cdot \mathbf{\sigma }\left( 
\begin{array}{c}
a_{\mathbf{k}} \\ 
b_{\mathbf{k}}%
\end{array}%
\right) ,
\end{equation}%
by Pauli matrices $\mathbf{\sigma }$, and then can be topologically
non-trivial or not, depending on the explicit form of $\mathbf{B}_{\mathbf{k}%
}$. ii) For a given $\mathbf{k}$, matrix $h_{\mathbf{k}}$\ contains a Jordan
block at instant $t=t_{\mathbf{k}}$, where $t_{\mathbf{k}}$ satisfies the
equation%
\begin{equation}
\mu (t_{\mathbf{k}})=\pm \left\vert \mathbf{B}(\mathbf{k)}\right\vert .
\end{equation}%
In this case, there are only two eigenstates for $h_{\mathbf{k}}$, which are
the eigenstates of $\mathbf{B}_{\mathbf{k}}\cdot \mathbf{\sigma }$, and then
the completeness of eigenstates is spoiled.

Actually, in general, matrix $h_{\mathbf{k}}$\ can be rewritten as 
\begin{equation}
h_{\mathbf{k}}=\left\vert B(\mathbf{k)}\right\vert \left( 
\begin{array}{ccc}
\cos \theta _{\mathbf{k}} & \sin \theta _{\mathbf{k}}e^{-i\varphi _{\mathbf{k%
}}} & 0 \\ 
\sin \theta _{\mathbf{k}}e^{i\varphi _{\mathbf{k}}} & -\cos \theta _{\mathbf{%
k}} & \gamma _{\mathbf{k}} \\ 
0 & 0 & \Delta _{\mathbf{k}}%
\end{array}%
\right),  \label{hk}
\end{equation}%
where parameters in the matrix elements are defined as 
\begin{eqnarray}
\tan \theta _{\mathbf{k}} &=&\frac{B_{z}(\mathbf{k)}}{\left\vert B(\mathbf{k)%
}\right\vert },\tan \varphi _{\mathbf{k}}=\frac{B_{x}(\mathbf{k)}}{B_{y}(%
\mathbf{k)}},  \notag \\
\gamma _{\mathbf{k}} &=&\gamma /\varepsilon _{\mathbf{k}},\Delta _{\mathbf{k}%
}=\mu /\varepsilon _{\mathbf{k}},\varepsilon _{\mathbf{k}}=\left\vert B(%
\mathbf{k)}\right\vert .
\end{eqnarray}%
Note that although matrix $h_{\mathbf{k}}$\ is non-Hermitian, its
eigenvalues are always real. Three eigenvectors are obtained as%
\begin{eqnarray}
\left\vert \psi _{\mathbf{k}}^{+}\right\rangle &=&\left( 
\begin{array}{c}
\cos \frac{\theta _{k}}{2} \\ 
e^{i\varphi _{k}}\sin \frac{\theta _{k}}{2} \\ 
0%
\end{array}%
\right) ,\left\vert \psi _{\mathbf{k}}^{\mathbf{-}}\right\rangle =\left( 
\begin{array}{c}
-\sin \frac{\theta _{k}}{2} \\ 
e^{i\varphi _{k}}\cos \frac{\theta _{k}}{2} \\ 
0%
\end{array}%
\right) ,  \notag \\
\left\vert \psi _{\mathbf{k}}^{\Delta }\right\rangle &=&\left( 
\begin{array}{c}
\gamma _{k}e^{-i\varphi _{k}}\sin \theta _{k} \\ 
\gamma _{k}\left( \Delta _{k}-\cos \theta _{k}\right) \\ 
\Delta _{k}^{2}-1%
\end{array}%
\right) ,
\end{eqnarray}%
with the eigenvalues 
\begin{equation}
\varepsilon _{\mathbf{k}}^{\pm }=\pm \varepsilon _{\mathbf{k}},\varepsilon _{%
\mathbf{k}}^{\Delta }=\Delta _{k}\varepsilon _{\mathbf{k}}.
\end{equation}%
It shows that when taking $\Delta _{\mathbf{k}}=\pm 1$, or at $t=t_{\mathbf{k%
}}$, matrix $h_{\mathbf{k}}$\ reaches at the EP. And we have $\left\vert
\psi _{\mathbf{k}}^{\Delta }\right\rangle =\left\vert \psi _{\mathbf{k}%
}^{\pm }\right\rangle $, i.e., the coalescing state appears, which is
crucial for the scheme in this work.

\section{EP dynamics and periodic driving}

\label{EP dynamics and periodic driving}

Based on the above analysis, the dynamics of $H$ is governed by the time
evolution operator%
\begin{equation}
U(t)=\exp (-iHt)=\prod_{\mathbf{k}}U_{\mathbf{k}}(t),  \label{U(t)}
\end{equation}%
where the time evolution operator in sub-space $\mathbf{k}$ has the form 
\begin{equation}
U_{\mathbf{k}}(t)=\mathcal{T}\exp [-i\int_{0}^{t}H_{\mathbf{k}}(t^{\prime })%
\text{\textrm{d}}t^{\prime }],  \label{Uk(t)}
\end{equation}%
with $\mathcal{T}$ being the time-order operator. We first consider the case
with slowly varying $H_{\mathbf{k}}(t)$\ (i.e., very small $\omega $). The
time evolution around the instant $t=t_{\mathbf{k}}$\ is crucial and can be
described approximately by the operator%
\begin{equation}
U_{\mathbf{k}}(t)\approx \exp [-iH_{\mathbf{k}}(t_{\mathbf{k}})t],
\end{equation}%
which obeys an exclusive EP dynamics.

Considering a time-independent $h_{\mathbf{k}}^{\mathrm{EP}}$\ at EP,%
\begin{equation}
h_{\mathbf{k}}^{\mathrm{EP}}=\varepsilon _{\mathbf{k}}\left( 
\begin{array}{ccc}
\cos \theta _{\mathbf{k}} & \sin \theta _{\mathbf{k}}e^{-i\varphi _{\mathbf{k%
}}} & 0 \\ 
\sin \theta _{\mathbf{k}}e^{i\varphi _{\mathbf{k}}} & -\cos \theta _{\mathbf{%
k}} & \gamma _{\mathbf{k}} \\ 
0 & 0 & -1%
\end{array}%
\right) ,
\end{equation}%
which contains a Jordan block, satisfying%
\begin{equation}
h_{\mathbf{k}}^{\mathrm{EP}}\left\vert \psi _{\mathbf{k}}^{\pm
}\right\rangle =\pm \varepsilon _{\mathbf{k}}\left\vert \psi _{\mathbf{k}%
}^{\pm }\right\rangle ,h_{\mathbf{k}}^{\mathrm{EP}}\left\vert \psi _{\mathbf{%
k}}^{\mathrm{a}}\right\rangle =-\varepsilon _{\mathbf{k}}\left\vert \psi _{%
\mathbf{k}}^{\mathrm{a}}\right\rangle .
\end{equation}%
Here vector \ 
\begin{equation}
\left\vert \psi _{\mathbf{k}}^{\mathrm{a}}\right\rangle =\left( 0,0,1\right)
^{\mathrm{T}}
\end{equation}%
is referred to as auxiliary vector, while $\left\vert \psi _{\mathbf{k}%
}^{-}\right\rangle $\ is the coalescing state.

Straightforward derivation shows that%
\begin{eqnarray}
&&\exp \left( -ih_{\mathbf{k}}^{\mathrm{EP}}t\right) =-\frac{\gamma
_{k}\varepsilon _{\mathbf{k}}t}{2}  \notag \\
&&\times \lbrack \sin \left( \varepsilon _{\mathbf{k}}t\right) +i(1+2\sin
^{2}\frac{\varepsilon _{\mathbf{k}}t}{2})]A  \notag \\
&&+\cos \left( \varepsilon _{\mathbf{k}}t\right) I_{3}-i\sin \left(
\varepsilon _{\mathbf{k}}t\right) \left( \frac{h_{\mathbf{k}}^{\mathrm{EP}}}{%
\varepsilon _{\mathbf{k}}}+\frac{\gamma _{k}A}{2}\right) ,
\end{eqnarray}%
where the matrix 
\begin{equation}
A=\left( \frac{h_{\mathbf{k}}^{\mathrm{EP}}}{\varepsilon _{\mathbf{k}}}%
\right) ^{2}-I_{3}=\left( 
\begin{array}{ccc}
0 & 0 & -\sin \frac{\theta _{k}}{2} \\ 
0 & 0 & e^{i\varphi _{k}}\cos \frac{\theta _{k}}{2} \\ 
0 & 0 & 0%
\end{array}%
\right) ,
\end{equation}%
satisfying%
\begin{equation}
A\left\vert \psi _{\mathbf{k}}^{\mathrm{a}}\right\rangle =\left\vert \psi _{%
\mathbf{k}}^{-}\right\rangle .
\end{equation}%
The time evolution operator contains terms with\ linear and periodic
functions of $t$. Then the time evolution for initial state $\left\vert \psi
_{\mathbf{k}}^{\mathrm{a}}\right\rangle $ is%
\begin{eqnarray}
&&\left\vert \psi _{\mathbf{k}}^{\mathrm{EP}}\left( t\right) \right\rangle
=\exp \left( -ih_{\mathbf{k}}^{\mathrm{EP}}t\right) \left\vert \psi _{%
\mathbf{k}}^{\mathrm{a}}\right\rangle  \notag \\
&=&-\frac{\gamma _{k}\varepsilon _{\mathbf{k}}t}{2}\times \lbrack \sin
\left( \varepsilon _{\mathbf{k}}t\right) +i(1+2\sin ^{2}\frac{\varepsilon _{%
\mathbf{k}}t}{2})]\left\vert \psi _{\mathbf{k}}^{-}\right\rangle  \notag \\
&&+\cos \left( \varepsilon _{\mathbf{k}}t\right) \left\vert \psi _{\mathbf{k}%
}^{\mathrm{a}}\right\rangle -i\sin \left( \varepsilon _{\mathbf{k}}t\right)
\left( \frac{\gamma _{k}}{2}-1\right) \left\vert \psi _{\mathbf{k}%
}^{-}\right\rangle .
\end{eqnarray}%
It indicates that the long-time evolution of initial state\ $d_{\mathbf{k}%
}^{\dag }\left\vert 0\right\rangle _{\mathbf{k}}$\ ($\left\vert
0\right\rangle _{\mathbf{k}}$ is the vacuum state for operators $a_{\mathbf{k%
}}$, $b_{\mathbf{k}}$, and $d_{\mathbf{k}}$.) turns to the coalescing state $%
\left\vert \psi _{\mathbf{k}}^{-}\right\rangle $ due to the linear-time
dependence of the first term.\ Obviously, the action of $U_{\mathbf{k}}(t)$\
at long time is the projection of any pure initial state on the component $%
\left\vert \psi _{\mathbf{k}}^{-}\right\rangle $, which is perfect transport
of fermion from the source to the central system. For many-particle case, we
note that all the time-dependent $h_{\mathbf{k}}$\ can not reach their\ $%
h_{k}^{\mathrm{EP}}$, simultaneously. The dynamics for $h_{k}$\textbf{\ }%
near EP may result in the oscillation of the particle number between the
source and the central systems, although the hopping term is unidirectional.
Nevertheless, periodically varying $\mu (t)$ passes by the EP of every $h_{%
\mathbf{k}}$ is expected to transport the fermion from source to the central
system in each $\mathbf{k}$ sector near completely.\textbf{\ }Ideally, if
this occurs in every $\mathbf{k}$ sector, the full-filled initial state

\begin{equation}
\prod_{j}d_{j}^{\dag }\left\vert 0\right\rangle =\prod_{\mathbf{k}}d_{%
\mathbf{k}}^{\dag }\left\vert 0\right\rangle
\end{equation}%
will evolve to the ground state of the central system meanwhile leaving
empty in the source system. Note that the initial state is a trivial
many-particle state, a saturatedly filled state. The aim of the scheme is to
prepare a nontrivial state from such an initial state through the time
evolution in the context of non-Hermitian quantum mechanics.

\begin{figure}[tb]
\centering
\includegraphics[ bb=290 0 850 540, width=0.23\textwidth, clip]{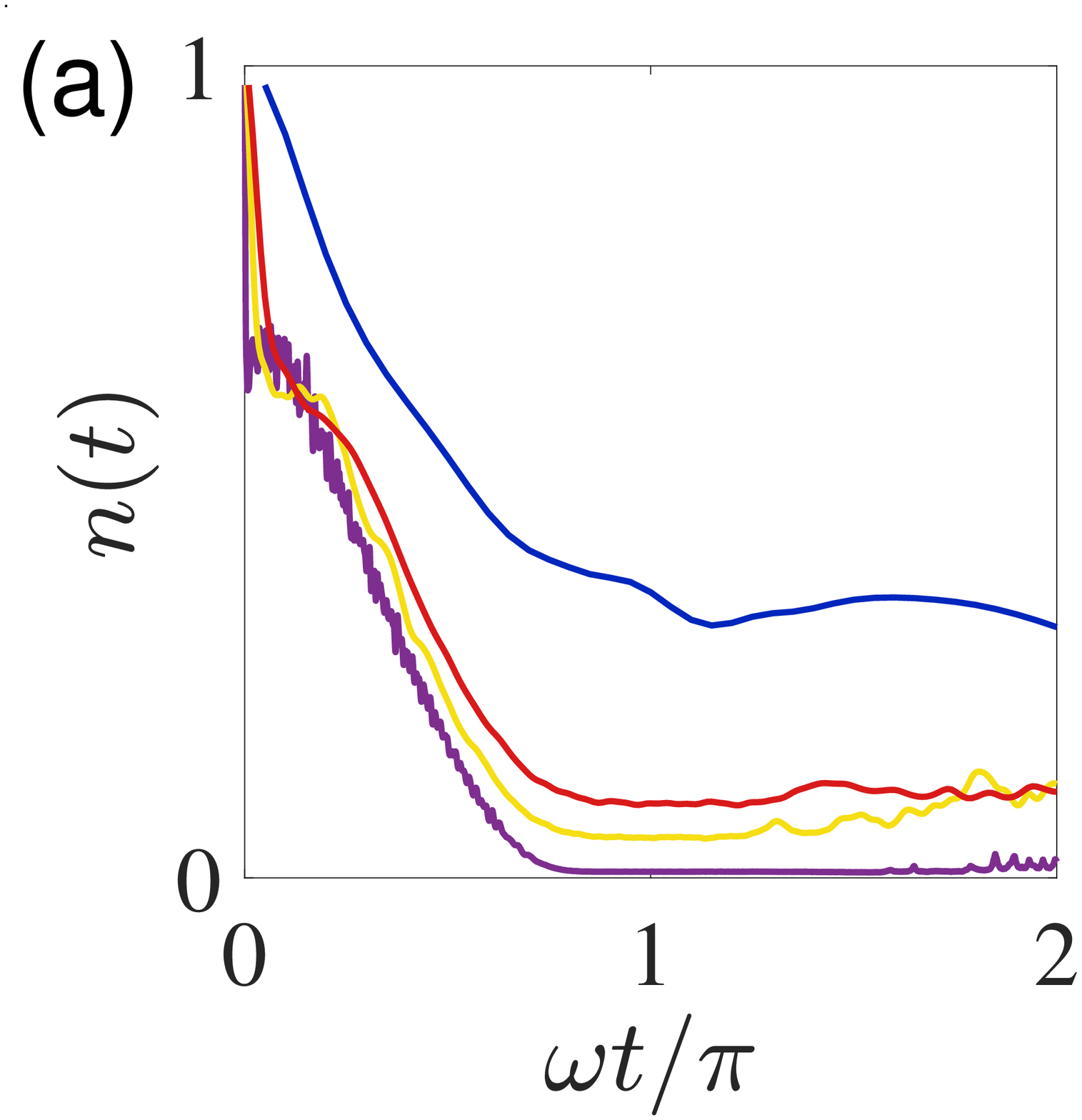} %
\includegraphics[ bb=290 0 850 540, width=0.23\textwidth, clip]{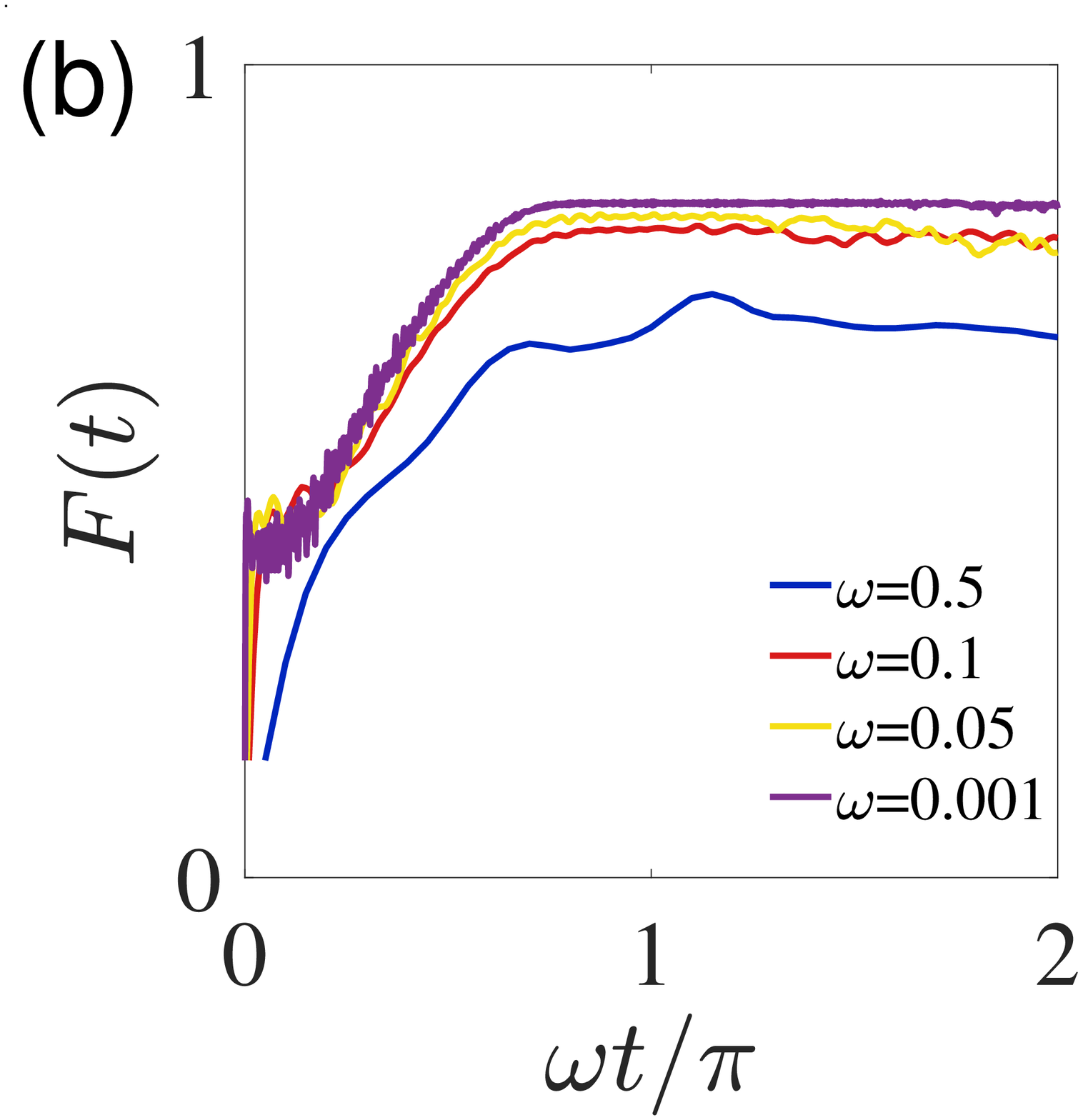}
\caption{Plots of $n\left( t\right) $ and $F\left( t\right) $, which are
defined in Eqs. (\protect\ref{nm}) and Eq. (\protect\ref{fm}), obtained by
exact diagonalization for finite system. The parameters are $N=20\times 20$, 
$u=1.2$, $\protect\mu _{0}=-2$, $W=2.6$, and $\protect\gamma =0.5$. Four
typical values of $\protect\omega $ are taken, and indicated in the panels.}
\label{fig2}
\end{figure}

To illustrate our scheme and investigate its efficiency, we consider the
central system as QWZ model introduced by Qi, Wu and Zhang \cite{QWZ}. The
Bloch Hamiltonian is

\begin{equation}
h_{\mathbf{k}}=B_{x}\sigma _{x}+B_{y}\sigma _{y}+B_{z}\sigma _{z},
\label{QWZ}
\end{equation}%
where the field components are%
\begin{eqnarray}
B_{x} &=&\sin k_{x},B_{y}=\sin k_{y},  \notag \\
B_{z} &=&u+\cos k_{x}+\cos k_{y}.
\end{eqnarray}%
It is well known that the Chern number of the lower band is 
\begin{eqnarray}
c &=&0,\text{for }\left\vert u\right\vert >2,  \notag \\
c &=&\pm 1,\text{for }0<\pm u<2.
\end{eqnarray}%
Numerical simulation is performed to verify the efficiency of the scheme.
Here the computation for time-ordered integral\ is performed by using a
uniform mesh in the time discretization for the time-dependent Hamiltonian $%
h_{\mathbf{k}}(t)$ with deferent $\omega $. We compute the time evolution
for the initial state $\left\vert \psi _{\mathbf{k}}\left( 0\right)
\right\rangle =d_{\mathbf{k}}^{\dag }\left\vert 0\right\rangle _{\mathbf{k}}$%
, and compare the evolved state $\left\vert \psi _{\mathbf{k}}\left(
t\right) \right\rangle $ with the target state%
\begin{equation}
\left\vert \psi _{\mathbf{k}}^{\text{c}}\right\rangle =\left( \sin \frac{%
\theta _{k}}{2}a_{\mathbf{k}}^{\dag }-e^{i\varphi _{k}}\cos \frac{\theta _{k}%
}{2}b_{\mathbf{k}}^{\dag }\right) \left\vert 0\right\rangle _{\mathbf{k}},
\end{equation}%
which is the coalescing state of $h_{\mathbf{k}}$ [Eq. (\ref{QWZ})] at EP
with negative energy. The observables are particle number left in the source
system%
\begin{equation}
n\left( t\right) =\frac{1}{N}\sum_{\mathbf{k}}\frac{\left\langle \psi _{%
\mathbf{k}}\left( t\right) \right\vert d_{\mathbf{k}}^{\dag }d_{\mathbf{k}%
}\left\vert \psi _{\mathbf{k}}\left( t\right) \right\rangle }{\left\vert
\left\vert \psi _{\mathbf{k}}\left( t\right) \right\rangle \right\vert ^{2}},
\label{nm}
\end{equation}%
and the fidelity%
\begin{equation}
F\left( t\right) =\frac{1}{N}\sum_{\mathbf{k}}\frac{\left\vert \langle \psi
_{\mathbf{k}}^{\text{c}}\left\vert \psi _{\mathbf{k}}\left( t\right)
\right\rangle \right\vert }{\left\vert \left\vert \psi _{\mathbf{k}}^{\text{c%
}}\right\rangle \left\vert \psi _{\mathbf{k}}\left( t\right) \right\rangle
\right\vert },  \label{fm}
\end{equation}%
which measure the efficiency of the transport. The numerical results for
finite systems with presentative parameters are plotted in Fig. \ref{fig2}.
We see that the optimal efficiency occurs when $\omega =0.001$, and the
transport efficiency decreases gradually with the increasing of $\omega .$

\begin{figure*}[tb]
\centering
\includegraphics[bb=163 0 1050 540, width=0.32\textwidth, clip]{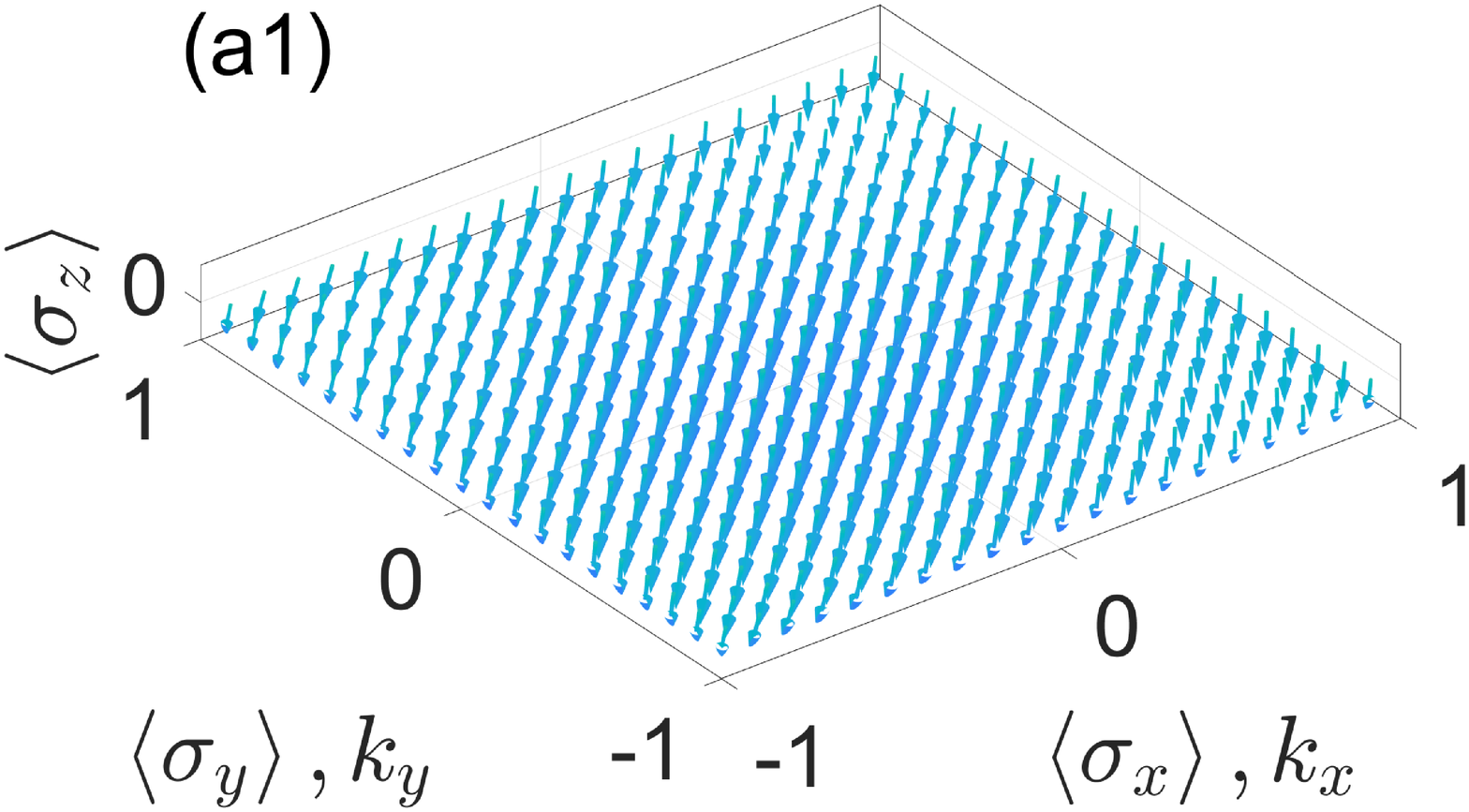} %
\includegraphics[bb=163 0 1050 540, width=0.32\textwidth, clip]{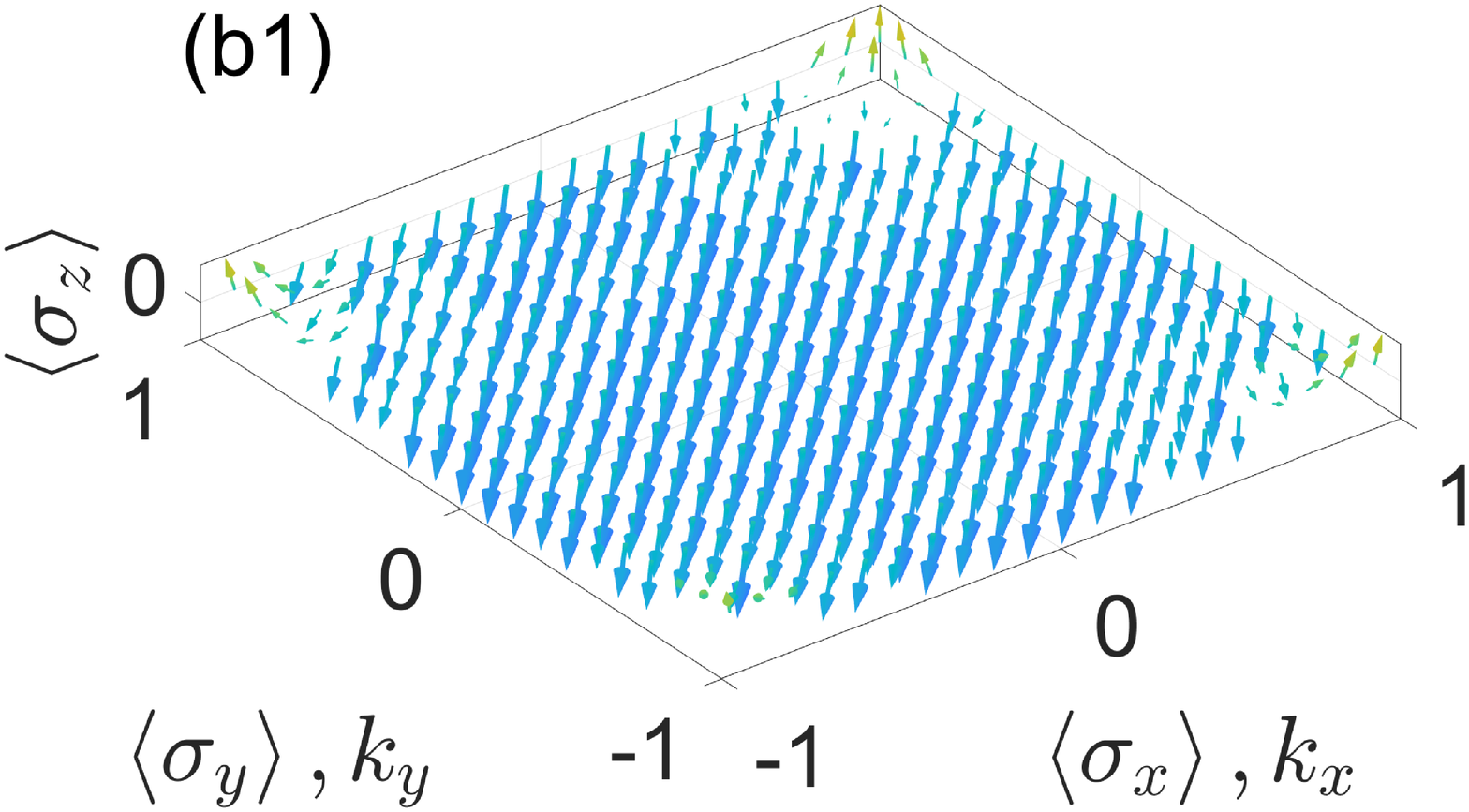} %
\includegraphics[bb=163 0 1050 540, width=0.32\textwidth, clip]{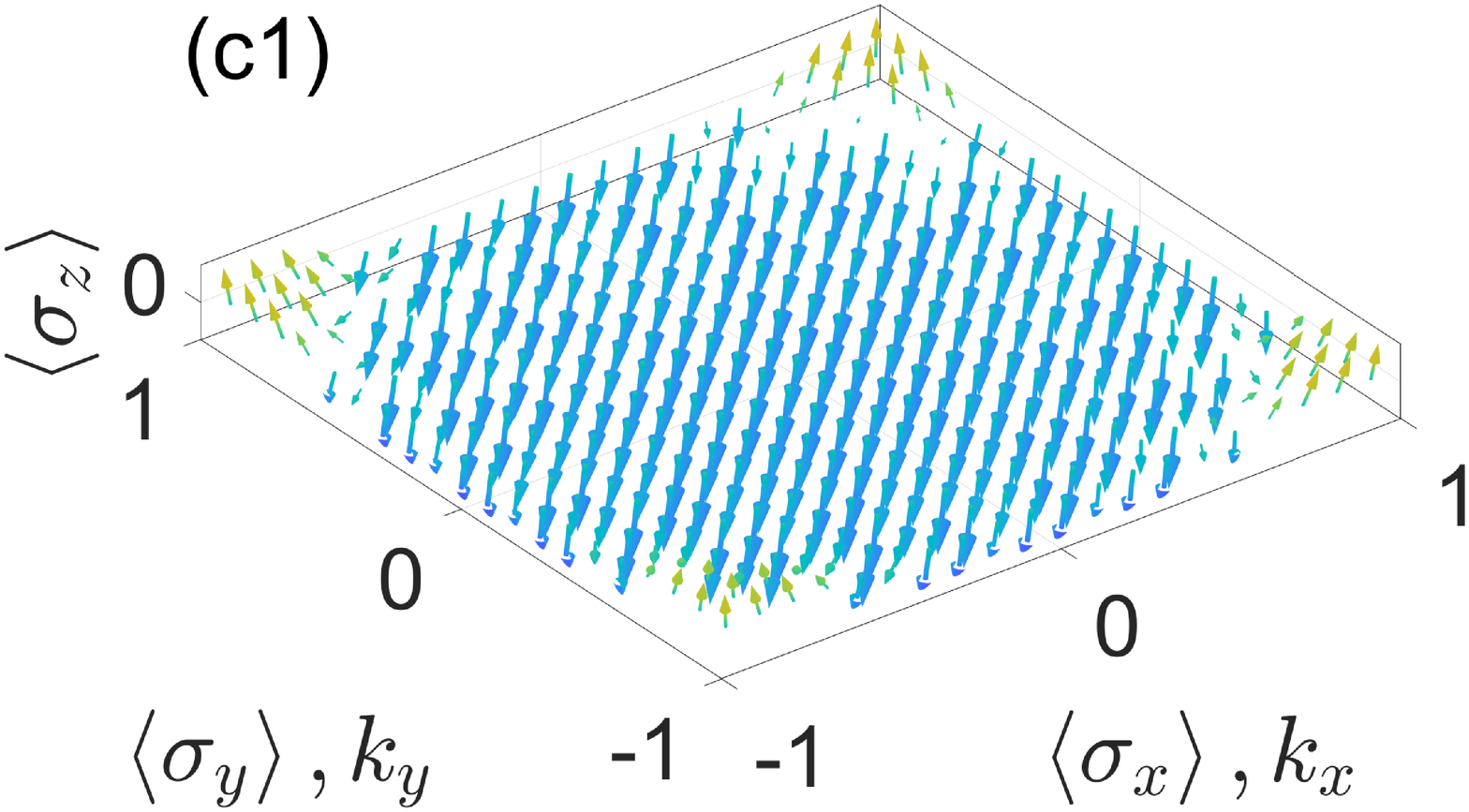} %
\includegraphics[bb=163 0 1050 540, width=0.32\textwidth, clip]{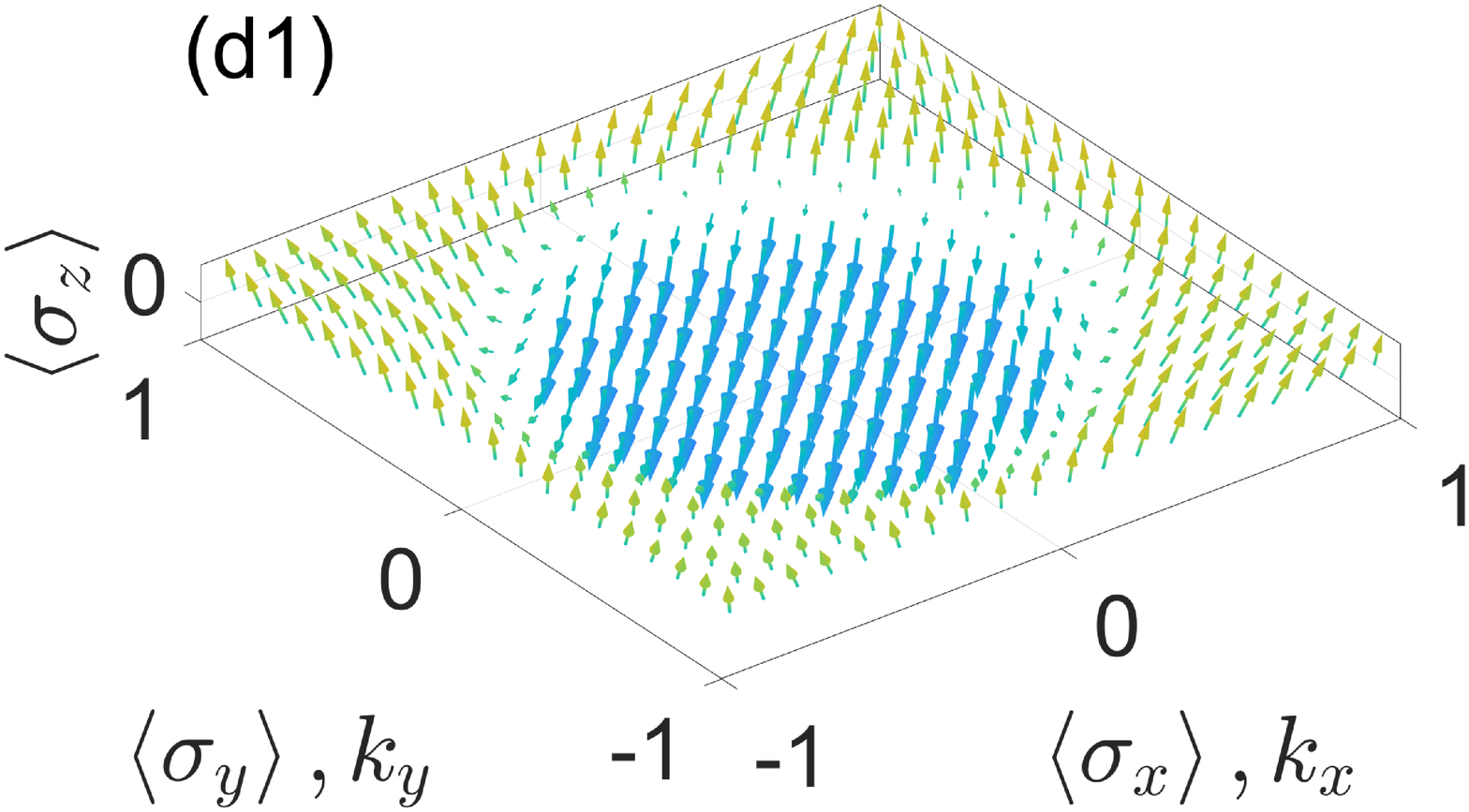} %
\includegraphics[bb=163 0 1050 540, width=0.32\textwidth, clip]{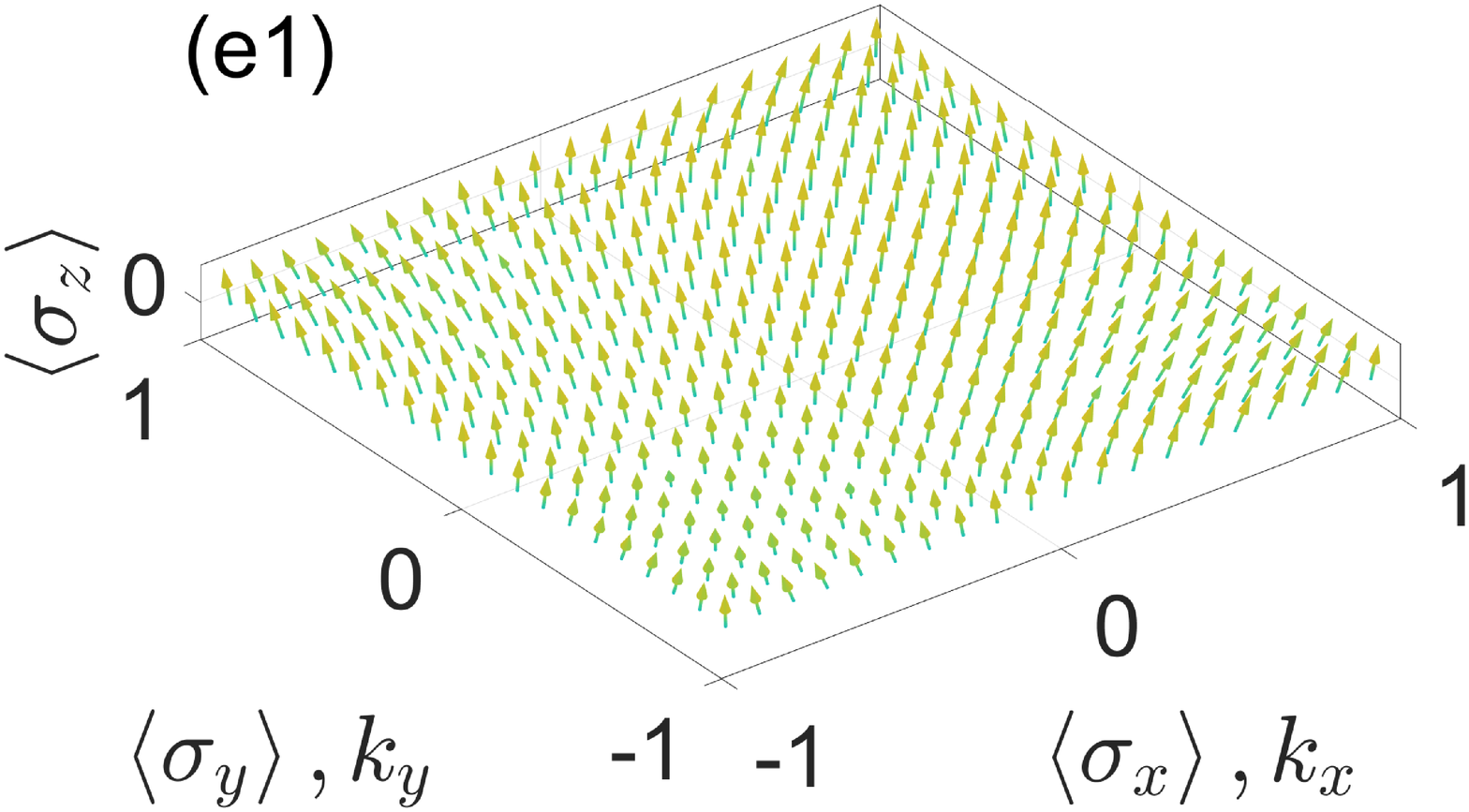} %
\includegraphics[bb=163 10 1050 600, width=0.3\textwidth, clip]{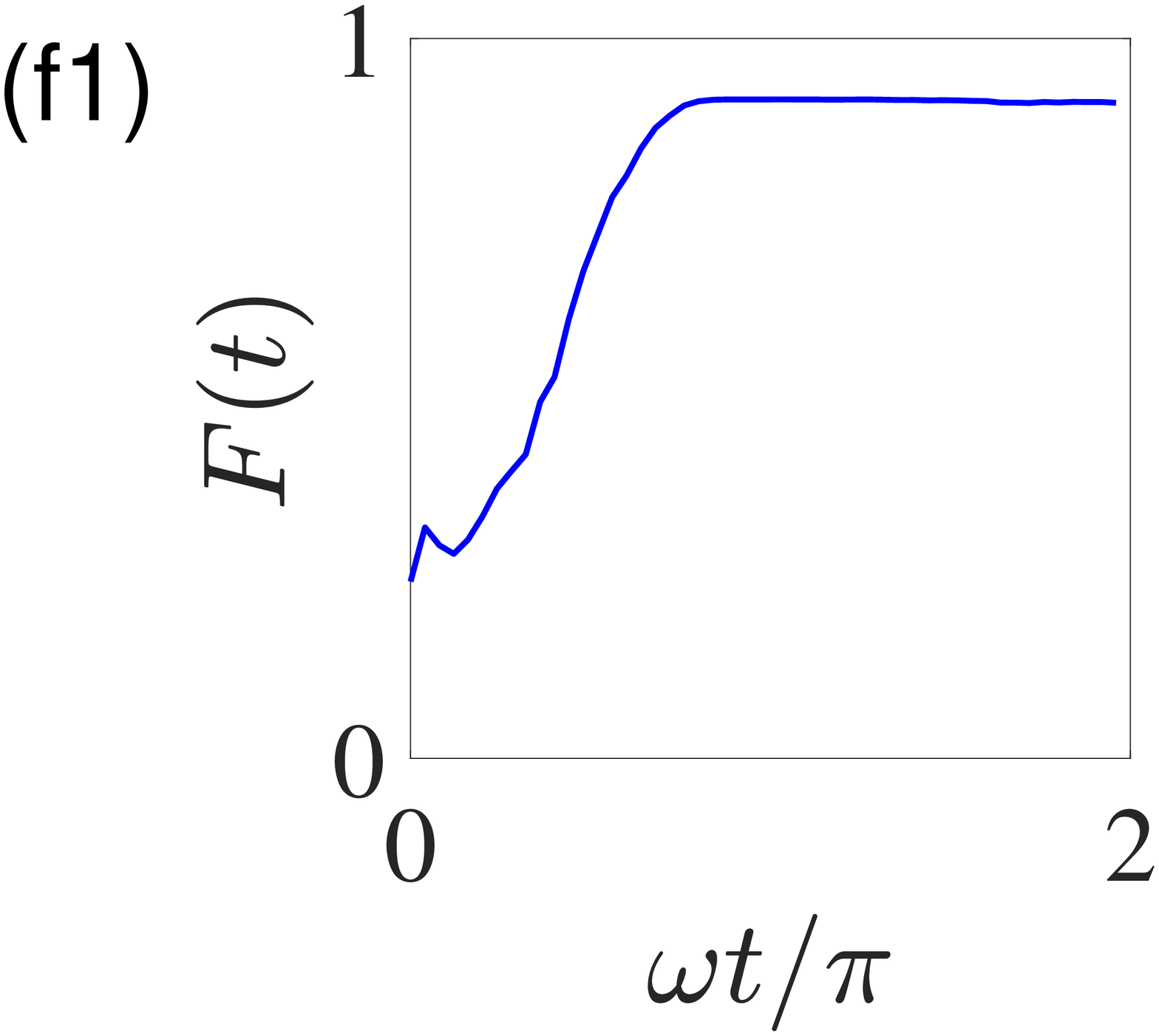} %
\includegraphics[bb=163 0 1050 540, width=0.32\textwidth, clip]{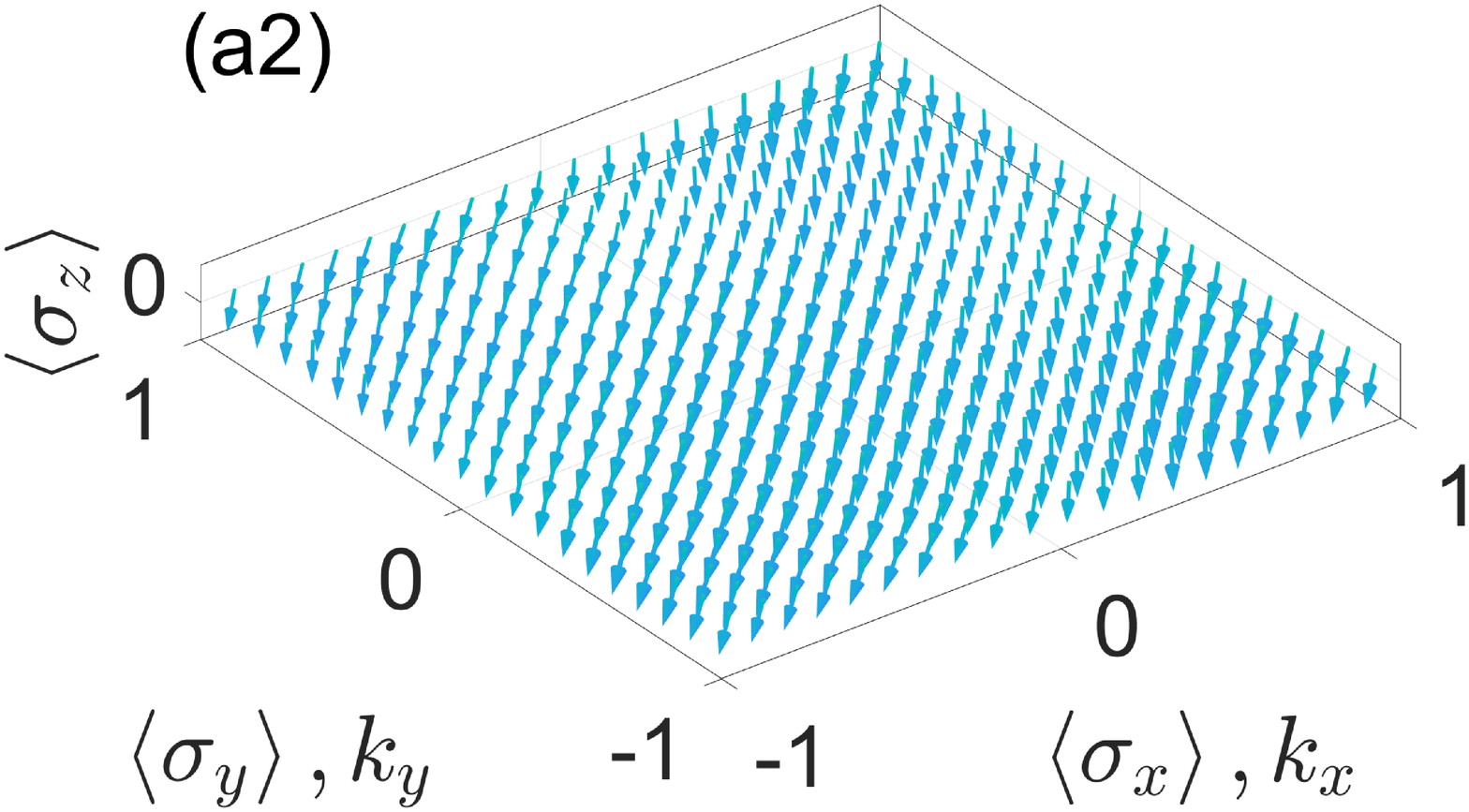} %
\includegraphics[bb=163 0 1050 540, width=0.32\textwidth, clip]{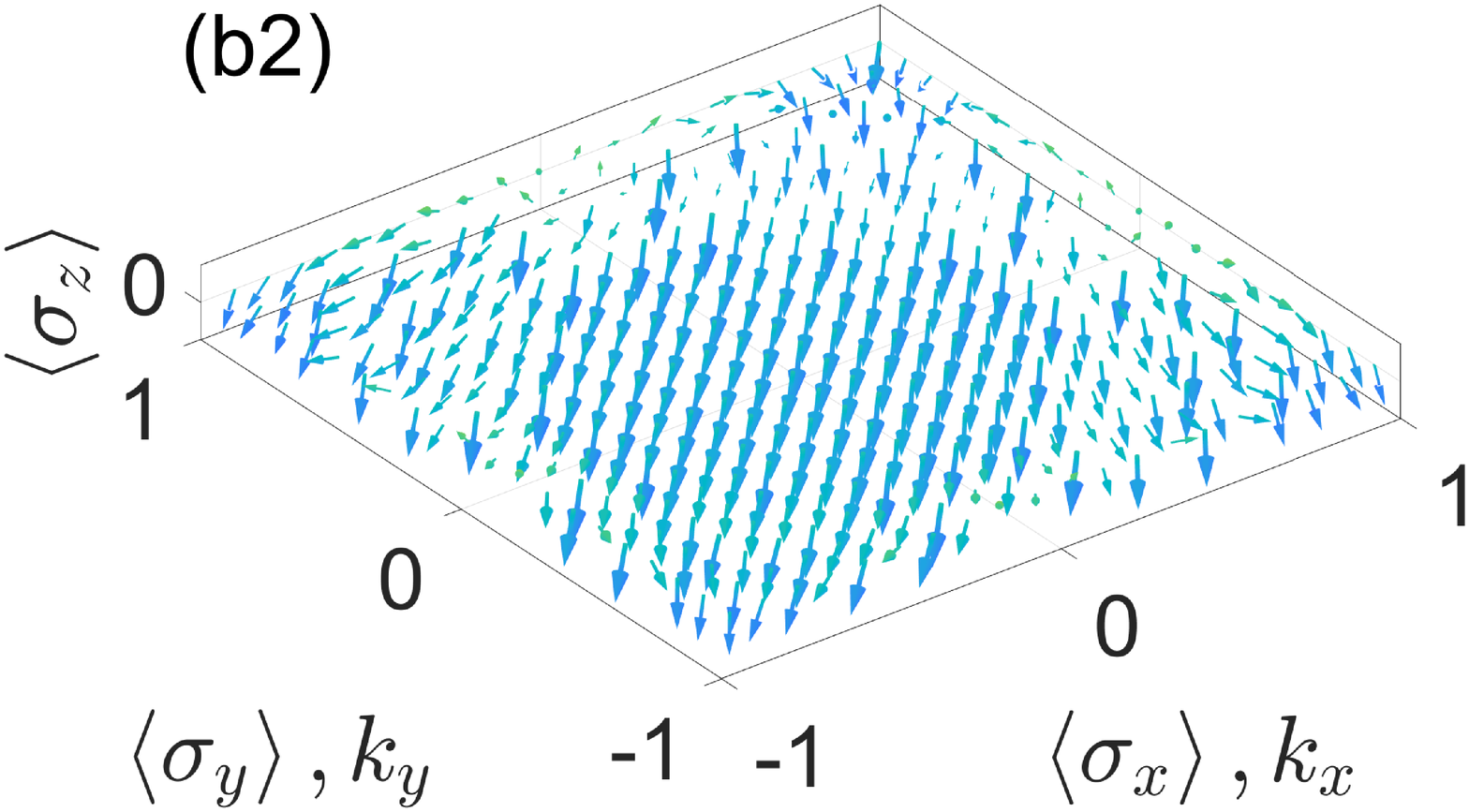} %
\includegraphics[bb=163 0 1050 540, width=0.32\textwidth, clip]{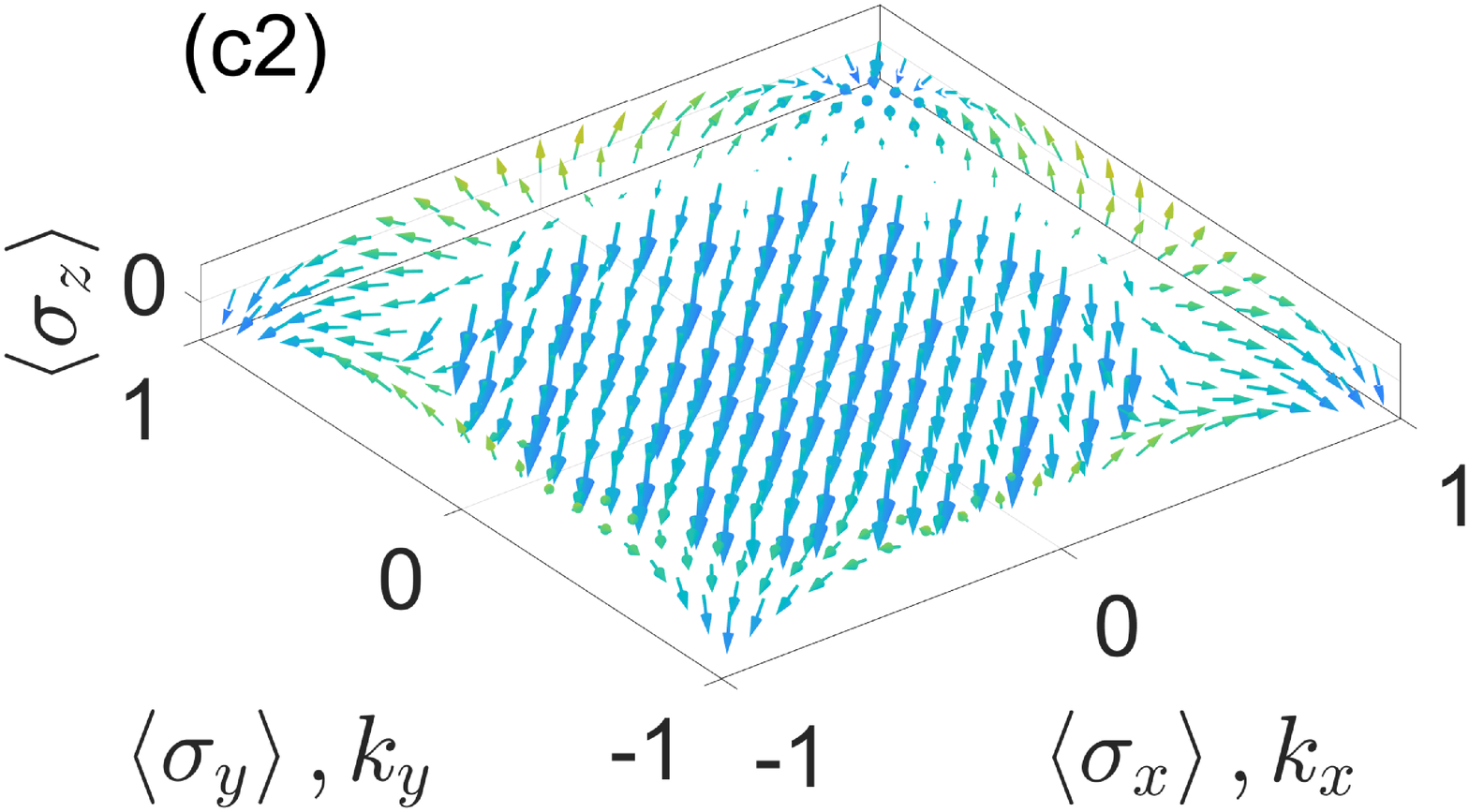} %
\includegraphics[bb=163 0 1050 540, width=0.32\textwidth, clip]{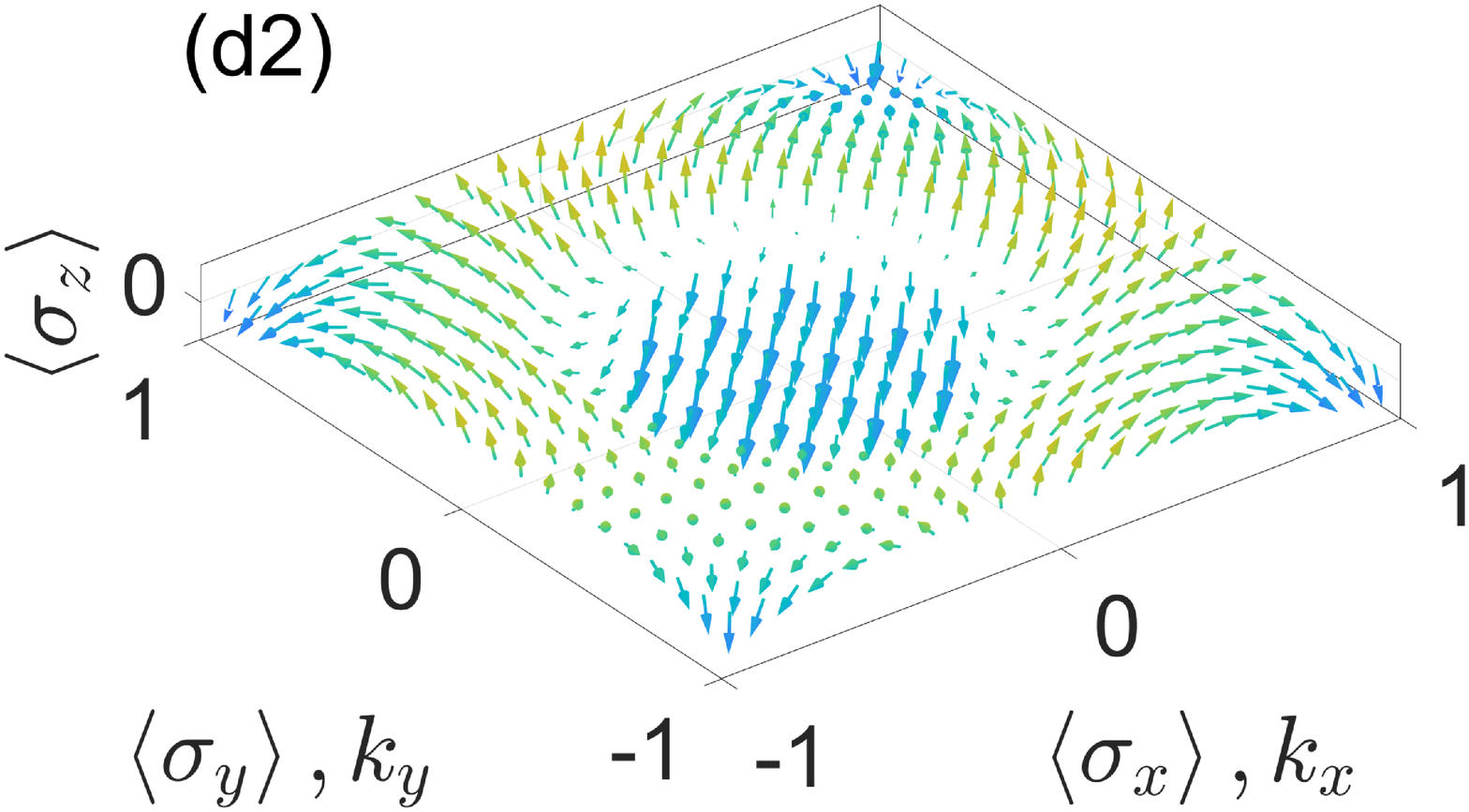} %
\includegraphics[bb=163 0 1050 540, width=0.32\textwidth, clip]{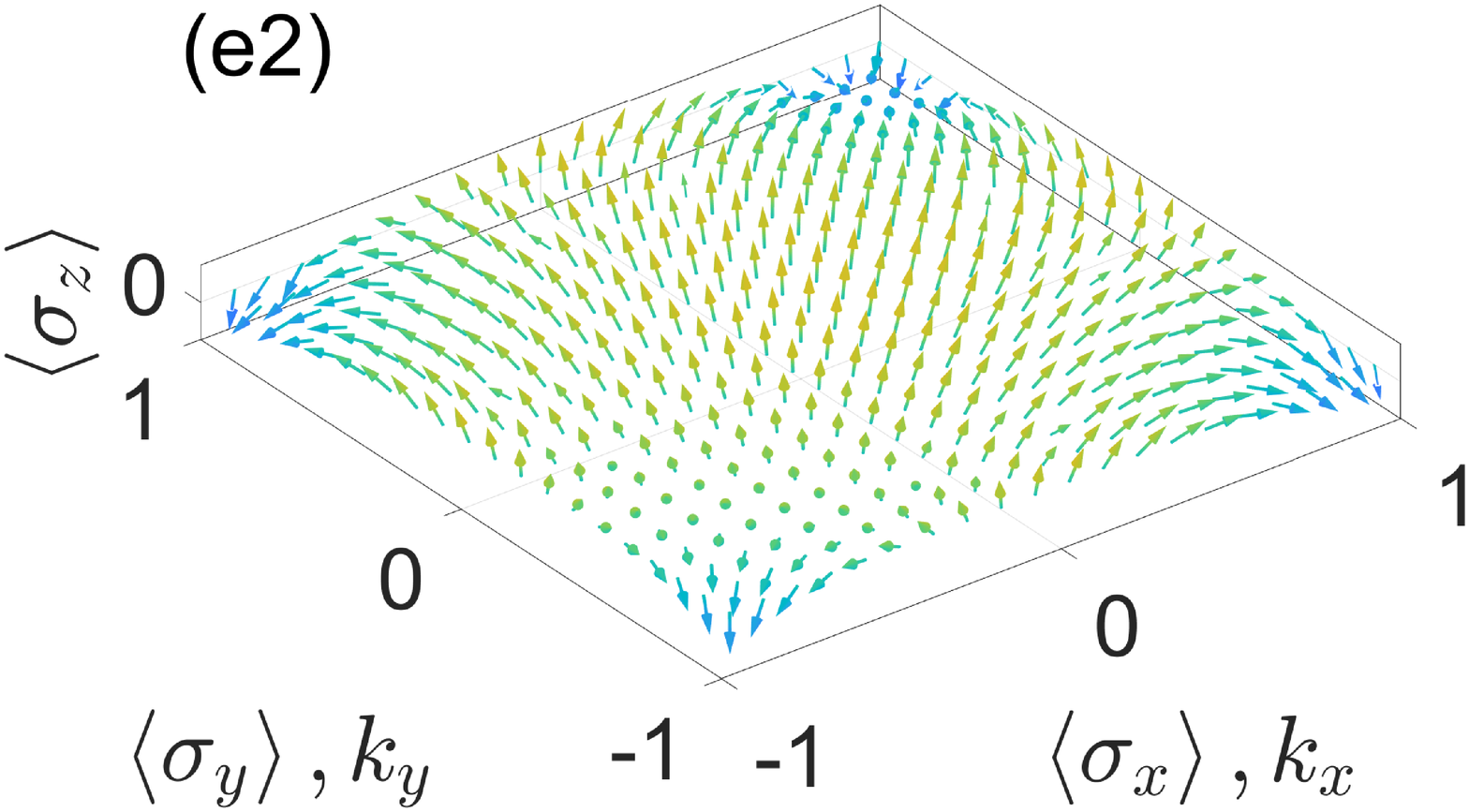} %
\includegraphics[bb=163 10 1050 600, width=0.3\textwidth, clip]{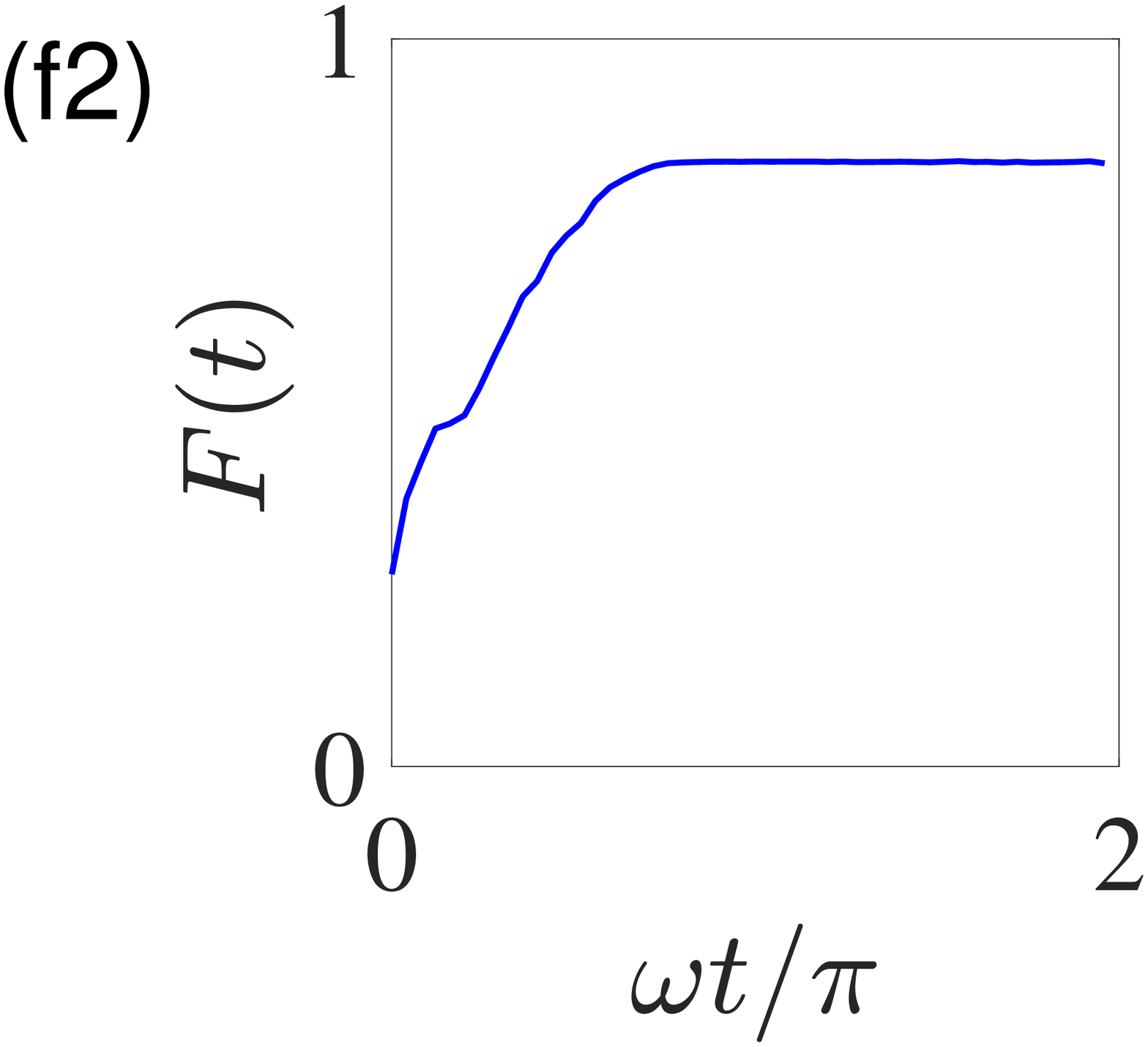}
\caption{Plots of skyrmion pattern at several typical instants, which are
defined in Eq. (\protect\ref{skyr}), obtained by exact diagonalization for
finite system. The corresponding fidelity is also plotted as comparison. The
parameters are $N=20\times 20$, $\protect\omega =0.001$, $\protect\gamma %
=0.5 $, (a1)-(d1) $u=3.2$, $\protect\mu _{0}=-3.21$, $W=3.94$, and (a2)-(d2) 
$u=1.2$, $\protect\mu _{0}=-2$, $W=2.38$, which correspond to the central
systems in topologically trivial and non-trivial phases, respectively. }
\label{fig3}
\end{figure*}

We also compute the time-dependent skyrmion to characterize the formation
process of the target state. To this end we introduce the auxiliary matrices%
\begin{equation}
\Sigma _{\alpha }=\left( 
\begin{array}{cc}
\sigma _{\alpha } & 0 \\ 
0 & 1%
\end{array}%
\right) ,\Sigma _{0}=\left( 
\begin{array}{cc}
I_{2} & 0 \\ 
0 & 1%
\end{array}%
\right) ,
\end{equation}%
where $\sigma _{\alpha }$\ is Pauli matrix in $\alpha $ ($\alpha =x,y,z$)\
component and $I_{2}$\ is unit matrix. The time-dependent skyrmion is
evaluated as%
\begin{equation}
\left\langle \sigma _{\alpha }\right\rangle _{\mathbf{k},t}=\frac{%
\left\langle \psi _{\mathbf{k}}\left( t\right) \right\vert \Sigma _{\alpha
}\left\vert \psi _{\mathbf{k}}\left( t\right) \right\rangle }{\left\langle
\psi _{\mathbf{k}}\left( t\right) \right\vert \Sigma _{0}\left\vert \psi _{%
\mathbf{k}}\left( t\right) \right\rangle }.  \label{skyr}
\end{equation}%
Unlike the expectation value of Pauli matrix $\sigma _{\alpha }$\ in the
usual study \cite{LSPRA}, $\left\langle \sigma _{\alpha }\right\rangle _{%
\mathbf{k},0}=0$\ for all $\alpha $\ and $\sum_{\alpha =x,y,z}\left(
\left\langle \sigma _{\alpha }\right\rangle _{\mathbf{k},t}\right)
^{2}\leqslant 1$.\ In the case that all the fermions have been transported
to the central system in long-time limit the skyrmion obeys the pattern%
\begin{equation}
\left\langle \mathbf{\sigma }\right\rangle _{\mathbf{k},\infty }=\frac{%
\mathbf{B}(\mathbf{k)}}{\left\vert \mathbf{B}(\mathbf{k)}\right\vert },
\label{skyrmion}
\end{equation}%
which characterizes the topological feature of the phase. The numerical
results for finite systems with presentative parameters, $\left\langle
\sigma _{\alpha }\right\rangle _{\mathbf{k},t}$ at different time are
plotted in Fig. \ref{fig3}. Fig. \ref{fig3} (e1) and (e2) clearly show that
in long-time limit the skyrmion exhibits approximately the pattern defined
in Eq. (\ref{skyrmion}), which correspond to the central systems in
topologically trivial and non-trivial phases, respectively.

\section{Edge states engineering}

\label{Edge states engineering}

The aforementioned formalism is developed in the system with translational
symmetry in order to simplify the calculation procedure. This section will
be devoted to the realization of quantum mold casting in a system without
the translational symmetry. The essential step for this extension is
replacing the index $\mathbf{k}$\ by the eigenmodes of $H_{\mathrm{c}}$. Two
facts, flat band of $H_{\mathrm{s}}$\ and uniform hopping in $H_{\mathrm{in}%
} $, still allow $H$ to be block diagonalizable.

To demonstrate this point, we consider the central system $H_{\mathrm{c}}$\
to be a generalized RM chain with the Hamiltonian

\begin{eqnarray}
H_{\mathrm{c}} &=&\sum_{j=1}^{N-1}\left( w_{j}a_{j}^{\dag
}b_{j}+v_{j}a_{j+1}^{\dag }b_{j}+\mathrm{H.c.}\right)  \notag \\
&&+V\sum_{j=1}^{N}\left( a_{j}^{\dag }a_{j}-b_{j}^{\dag }b_{j}\right) ,
\label{RMC}
\end{eqnarray}%
where $w_{j}$\ and $v_{j}$ are position dependent hopping amplitudes
(including random distributions). The other Hamiltonians has a slight change
from the original form\ 
\begin{equation}
H_{\mathrm{s}}=\mu (t)\sum_{j=1}^{N}d_{j}^{\dag }d_{j}\text{, }H_{\mathrm{in}%
}=\gamma \sum_{j=1}^{N}\left( a_{j}^{\dag }d_{j}+b_{j}^{\dag }d_{j}\right) .
\label{RMS}
\end{equation}%
\begin{figure}[tb]
\centering
\includegraphics[ bb=120 550 300 750, width=0.27\textwidth, clip]{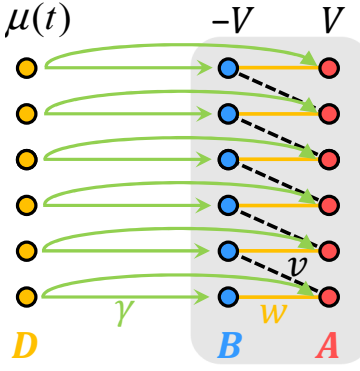}
\caption{schematic of the system with a generalized RM chain as the central
system $H_{\text{c}}$. Each part of the system is represented by Eq.(\protect
\ref{RMC}) and Eq.(\protect\ref{RMS}), respectively.}
\label{fig4}
\end{figure}
A schematics of the system is presented in Fig. \ref{fig4}. The Hamiltonian $%
H_{\mathrm{c}}$\ can be written as the diagonal-block form 
\begin{eqnarray}
H &=&\sum_{n=1}^{N}H_{n}, \\
H_{n} &=&\left( 
\begin{array}{ccc}
f_{a,n}^{\dag } & f_{b,n}^{\dag } & f_{d,n}^{\dag }%
\end{array}%
\right) h_{n}\left( 
\begin{array}{c}
f_{a,n} \\ 
f_{b,n} \\ 
f_{d,n}%
\end{array}%
\right) ,
\end{eqnarray}%
which reduces the eigenproblem of the present $H$\ to that of $3\times 3$\
matrix. Here the Bloch-like matrix $h_{n}$\ has the form

\begin{equation}
h_{n}=\left( 
\begin{array}{ccc}
V & \varepsilon _{0}(n) & \gamma \\ 
\varepsilon _{0}(n) & -V & \gamma \\ 
0 & 0 & \mu (t)%
\end{array}%
\right) ,  \label{hn}
\end{equation}%
and three sets of canonical fermion operators are defined as%
\begin{eqnarray}
f_{a,n}^{\dag } &=&\sum_{j=1}^{N}A_{j}^{n}a_{j}^{\dag },\text{ }%
f_{b,n}^{\dag }=\sum_{j=1}^{N}B_{j}^{n}b_{j}^{\dag },  \notag \\
f_{d,n}^{\dag } &=&\sum_{j=1}^{N}B_{j}^{n}d_{j}^{\dag }.
\end{eqnarray}%
with real coefficients $A_{j}^{n}$\ and $B_{j}^{n}\ $being obtained by
single-particle eigenstates of $H_{\mathrm{c}}$\ at $V=0$, with eigenvalues $%
\pm \varepsilon _{0}(n)$, satisfying the orthonormal complete relations

\begin{eqnarray}
\sum_{j}\left( A_{j}^{m}\right) ^{\ast }A_{j}^{n} &=&\sum_{j}\left(
B_{j}^{m}\right) ^{\ast }B_{j}^{n}=\delta _{mn},  \notag \\
\sum_{n}\left( A_{i}^{n}\right) ^{\ast }A_{j}^{n} &=&\sum_{n}\left(
B_{i}^{n}\right) ^{\ast }B_{j}^{n}=\delta _{ij},
\end{eqnarray}%
Actually, the Hamiltonian of SSH chain is diagonalized as%
\begin{equation}
H_{\mathrm{c}}\left( V=0\right) =\sum_{n=1}^{N}\varepsilon
_{0}(n)(f_{+,n}^{\dag }f_{+,n}-f_{-,n}^{\dag }f_{-,n}),
\end{equation}%
where $\varepsilon _{0}(n)>0$ is the positive energy spectrum with $n\in
\lbrack 1,N]$, and 
\begin{equation}
f_{\pm ,n}^{\dag }=\frac{1}{\sqrt{2}}\left( f_{a,n}^{\dag }\pm f_{b,n}^{\dag
}\right) ,
\end{equation}%
due to the fact that SSH chain\ is a bipartite lattice. We note that $h_{n}$%
\ is essentially the counterpart of $h_{\mathbf{k}}$\ in the Eq. (\ref{hk})
with a slight difference. The time evolution driven by $h_{n}$\ is similar
to that of $h_{\mathbf{k}}$, including the EP dynamics.

The central system $H_{\mathrm{c}}$ is the simplest prototype of a
topologically nontrivial band insulator with a symmetry protected
topological phase \cite{Ryu,Wen}. In recent years, it has been attracted
much attention and extensive studies have been demonstrated \cite%
{XD,Hasan,Delplace,ChenS1,ChenS2,LS PRA,WRPRA,WRPRB}. In the uniform case, $%
w=w_{j}<v=v_{j}$, there are two edge sates with eigenvalues $\pm V$, for
large $N$, which are explicitly expressed as%
\begin{eqnarray}
\left\vert \text{L}\right\rangle &=&\Omega \sum\limits_{j=1}^{N}\left( -%
\frac{w}{v}\right) ^{j-1}a_{j}^{\dagger }\left\vert 0\right\rangle , \\
\left\vert \text{R}\right\rangle &=&\Omega \sum\limits_{j=1}^{N}\left( -%
\frac{w}{v}\right) ^{N-j}b_{j}^{\dagger }\left\vert 0\right\rangle ,
\end{eqnarray}%
where the normalization factor is $\Omega =\sqrt{1-\left( \frac{w}{v}\right)
^{2}}$. In addition, small random perturbations on $w$\ and $v$\ cannot
remove the edge states and change their eigenvalues. Taking suitable value
of $V$, two edge states can lie within the gap of the spectrum. In the
following, we perform numerical simulation of time evolution for the
full-filled initial state

\begin{equation}
\left\vert \psi \left( 0\right) \right\rangle =\prod_{j}d_{j}^{\dag
}\left\vert 0\right\rangle =\prod_{n}d_{n}^{\dag }\left\vert 0\right\rangle ,
\end{equation}%
by taking several different values of $\omega $\ in $\mu (t)$. The evolved
state obeys%
\begin{equation}
\left\vert \psi \left( t\right) \right\rangle =\prod_{n}\left\vert \psi
_{n}\left( t\right) \right\rangle =U(t)\prod_{n}d_{n}^{\dag }\left\vert
0\right\rangle ,
\end{equation}%
where the time evolution operator has the similar form as Eq. (\ref{U(t)}),%
\begin{equation}
U(t)=\prod_{n}U_{n}(t),
\end{equation}%
and the time evolution operator in sub-space $n$ has the form 
\begin{equation}
U_{n}(t)=\mathcal{T}\exp [-i\int_{0}^{t}H_{n}(t^{\prime })\text{\textrm{d}}%
t^{\prime }].
\end{equation}%
The purpose of this process is the generation of single-particle edge state
in $H_{\mathrm{c}}$\ at levels $V=0$, which is isolated in the midgap.

\begin{figure}[tb]
\centering\includegraphics[ bb=0 200 400 560, width=0.48\textwidth,
clip]{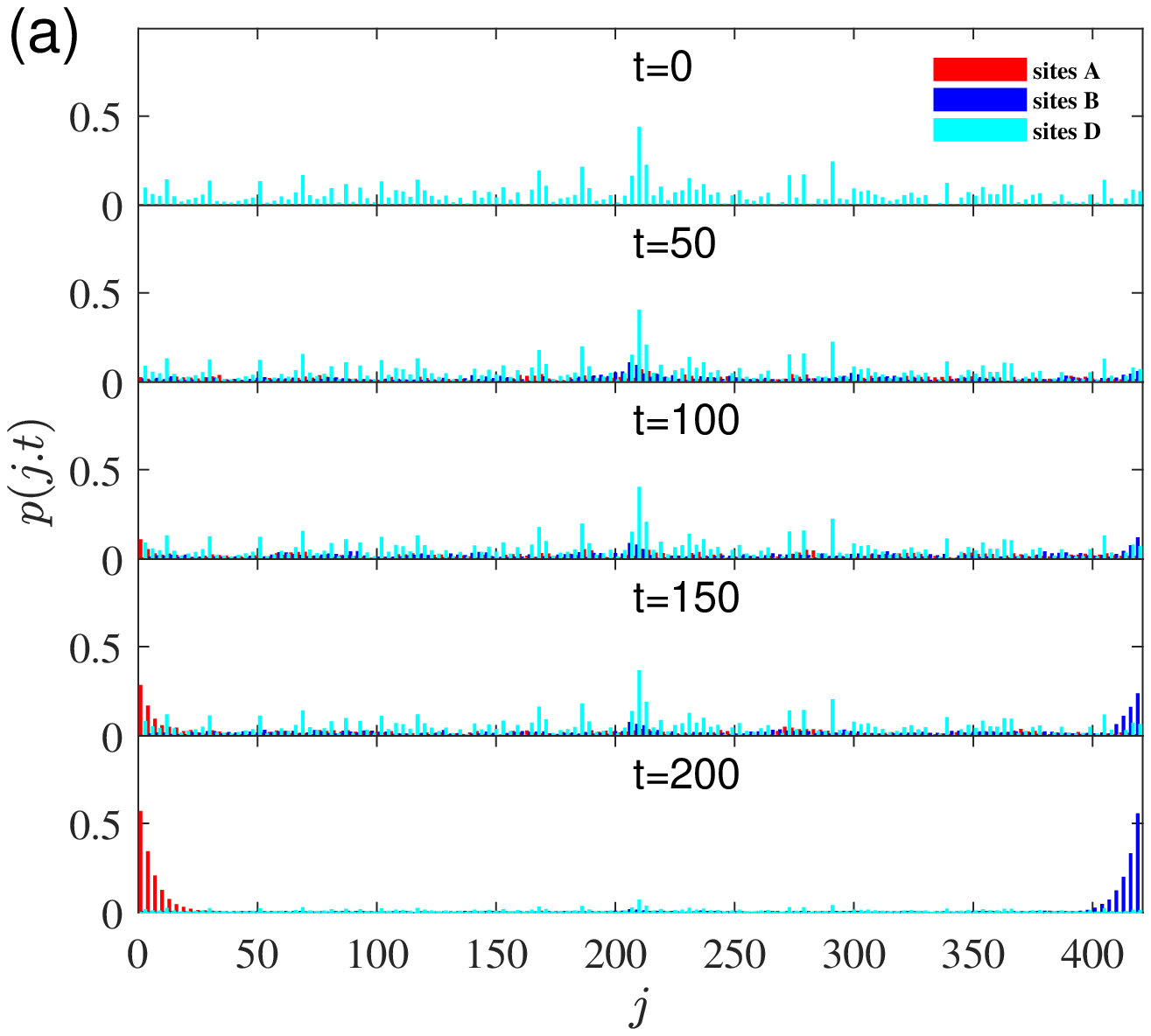} 
\includegraphics[ bb=276 220 850 534, width=0.48\textwidth,
clip]{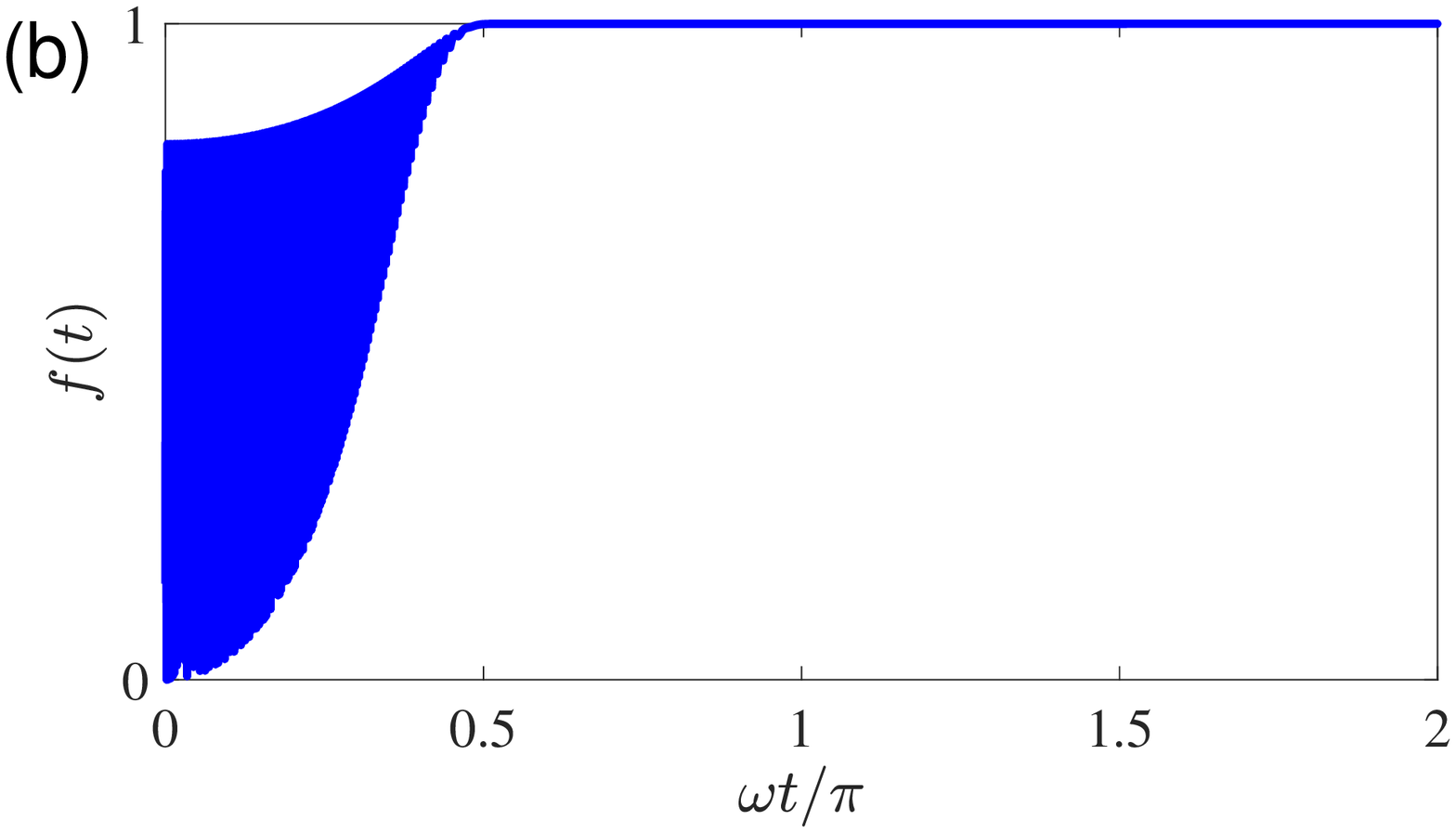}
\caption{(a) The profiles of $p\left( j,t\right) $ at several typical
instants, which is defined in Eq. (\protect\ref{plt}), showing the formation
of the edge state. (b) Corresponding fidelity defined in Eq. (\protect\ref%
{f(t)}). The parameters are $N=140,V=0,v=5,w=3,\protect\mu _{0}=0,W=2,%
\protect\gamma =0.5,\protect\omega =0.001.$}
\label{fig5}
\end{figure}

\begin{figure}[tb]
\centering%
\includegraphics[ bb=0 200 400 560, width=0.48\textwidth,
clip]{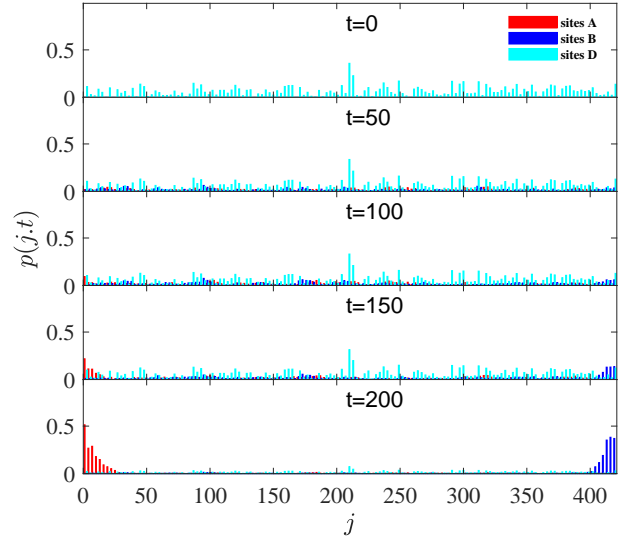}
\caption{The profiles of $p\left( j,t\right) $ at several typical instants
in the presence of random perturbations with different random strengths $%
w_{j}$ and $v_{j}$ defined in Eq. (\protect\ref{random}). The parameters are 
$N=140,V=0,v=5,w=3,\protect\mu _{0}=0,W=2,\protect\gamma =0.5,\protect\omega %
=0.001,R=1.2.$}
\label{fig6}
\end{figure}

We define the time-dependent distribution of particle probability\textbf{\ }$%
p\left( j,t\right) $\ in central system as%
\begin{eqnarray}
p\left( 3l-2,t\right) &=&\left\vert \left\langle 0\right\vert
a_{l}\left\vert \psi \left( t\right) \right\rangle \right\vert ,  \notag \\
p\left( 3l-1,t\right) &=&\left\vert \left\langle 0\right\vert
b_{l}\left\vert \psi \left( t\right) \right\rangle \right\vert ,  \notag \\
p\left( 3l,t\right) &=&\left\vert \left\langle 0\right\vert d_{l}\left\vert
\psi \left( t\right) \right\rangle \right\vert ,  \label{plt}
\end{eqnarray}%
for the evolved state to measure the efficiency of the scheme. Ideally, the
target state with perfect edge state has the distribution $p_{\mathrm{E}%
}\left( j\right) $ 
\begin{eqnarray}
p_{\mathrm{E}}\left( 3l-2\right) &=&\Omega ^{2}\left( \frac{w}{v}\right)
^{4l-2},  \notag \\
p_{\mathrm{E}}\left( 3l-1\right) &=&\Omega ^{2}\left( \frac{w}{v}\right)
^{2N-2l},  \notag \\
p_{\mathrm{E}}\left( 3l\right) &=&0.
\end{eqnarray}%
In the case with non-uniform distributions $\left\{ w_{j}\right\} $\ and $%
\left\{ v_{j}\right\} $, the corresponding $p_{\mathrm{E}}$\ can be obtained
numerically from exactly diagonalization of the Hamiltonian.\ Numerical
simulations for the formation processes of the single-particle edge state in
the absence and presence of random perturbations with different random
strengths.\textbf{\ }The computation is performed by taking two sets of
random numbers $\left\{ w_{j}\right\} $\ and $\left\{ v_{j}\right\} $ around 
$w$\ and $v$, i.e.,%
\begin{equation}
w_{j}=w+\mathrm{ran}(-R,R),v_{j}=v+\mathrm{ran}(-R,R),  \label{random}
\end{equation}%
where $\mathrm{ran}(-R,R)$\ denotes a uniform random number within $(-R,R)$.
We employ the fidelity%
\begin{equation}
f(t)=\left\vert \langle \psi _{\text{T}}\left\vert \psi \left( t\right)
\right\rangle \right\vert  \label{f(t)}
\end{equation}%
to characterize the efficiency of the scheme, where the target state $%
\left\vert \psi _{\text{T}}\right\rangle $ is the midgap state. The profiles
of $p\left( j,t\right) $\ for several representative situations with fixed $%
w $\ and $v$\ are presented in Fig. \ref{fig5} and with different random
strengths $\left\{ w_{j}\right\} $\ and $\left\{ v_{j}\right\} $\ are
presented in Fig. \ref{fig6}.\ We can see that the evolved state with fixed $%
w$\ and $v$\ very closes to the perfect edge state. In presence of random
perturbations, although the probability distribution seems irregular, it is
evidently edge state. These results accord with our predictions. This scheme
can be extended to the cases with $2$D and $3$D central systems for
preparing edge and surface states. Unlike the bulk states, these states are
responsible for the topological features.

\section{Summary}

\label{sec_summary}

In summary, we present a scheme to realize the quantum mold casting, i.e.,
engineering a target quantum state\ on demand by the time evolution of a
trivial initial state. The underlying mechanism is EP dynamics. We have
proposed a quantum mold model for dynamically casting a stable topological
insulating states and edge states. We introduce the periodic driving
chemical potential which causes EPs to exist in different subspaces, and
allows fermions be transferred from the full-filled trivial source system to
the corresponding subspaces of the topological central system. As examples,
we consider the central system as the QWZ model and generalized RM model to
dynamic casting topological insulating states and edge states, respectively.
Numerical simulations show that the scheme is efficient. The advantage of
the scheme is that the robust topological edge and surface states can be
engineered without filling the whole valence band.

\section*{Acknowledgement}

We acknowledge the support of NSFC (Grants No. 11874225).

\end{document}